\documentclass[aps,twocolumn,superscriptaddress]{revtex4}
\usepackage{graphicx}

\newcommand{\bra}[1] {\left\langle #1 \right|}
\newcommand{\ket}[1] {\left| #1 \right\rangle}

\begin{document}

\title{The information about the state of a charge qubit gained by a
weakly coupled quantum point contact}

%
%
%
%

\author{S. Ashhab}
\affiliation{Advanced Science Institute, The Institute of Physical
and Chemical Research (RIKEN), Wako-shi, Saitama 351-0198, Japan}
\affiliation{Physics Department, The University of Michigan, Ann
Arbor, Michigan 48109-1040, USA}

\author{J. Q. You}
\affiliation{Advanced Science Institute, The Institute of Physical
and Chemical Research (RIKEN), Wako-shi, Saitama 351-0198, Japan}
\affiliation{Department of Physics and Surface Physics Laboratory
(National Key Laboratory), Fudan University, Shanghai 200433,
China}

\author{Franco Nori}
\affiliation{Advanced Science Institute, The Institute of Physical
and Chemical Research (RIKEN), Wako-shi, Saitama 351-0198, Japan}
\affiliation{Physics Department, The University of Michigan, Ann
Arbor, Michigan 48109-1040, USA}

\begin{abstract}
We analyze the information that one can learn about the state of a
quantum two-level system, i.e.~a qubit, when probed weakly by a
nearby detector. We consider the general case where the qubit
Hamiltonian and the qubit's operator probed by the detector do not
commute. Because the qubit's state keeps evolving while being
probed and the measurement data is mixed with a detector-related
background noise, one might expect the detector to fail in this
case. We show, however, that under suitable conditions and by
proper analysis of the measurement data useful information about
the initial state of the qubit can be extracted. Our approach
complements the usual master-equation and quantum-trajectory
approaches, which describe the evolution of the qubit's quantum
state during the measurement process but do not keep track of the
acquired measurement information.
\end{abstract}

%

\maketitle

\section{Introduction}

Solid-state systems are among the most promising candidates for
future quantum information processing devices. One type of such
systems are superconductor- and semiconductor-based charge qubits
\cite{SolidStateReviews}. These qubits are commonly measured by
capacitively coupling them to quantum point contacts (QPC) or
single-electron transistors (SET), such that the current in the
detector is sensitive to the charge state of the qubit
\cite{Gurvitz,Korotkov,Makhlin,Goan,Pilgram,QuantumCapacitance,Johansson}.
By measuring the current passing through the detector, one can
infer the state of the qubit. One limitation that arises in
practical situations is that, in order to minimize the effects of
the detector on the qubit before the measurement, the
qubit-detector coupling strength is set to a value that is small
compared to the qubit's energy scale \cite{AshhabCoupling}. As a
result one must deal with some form of weak-measurement regime.
This type of weak, charge-sensitive readout works well when the
qubit is biased such that the charge states are eigenstates of the
Hamiltonian and therefore do not mix during the measurement. In
this case one can allow the detector to probe the qubit for as
long as is needed to obtain a high signal-to-noise ratio, without
having to worry about any coherent qubit dynamics (Note that,
since we are mainly interested in the measurement process, we
ignore any additional qubit decoherence mechanisms in the system,
which could impose constraints on the allowed measurement time).

In contrast to the simple situation described above, when the
detector weakly probes the charge state of the qubit while the
Hamiltonian induces mixing dynamics between the different charge
states, it becomes unclear how to interpret a given measurement
signal. Since the signal typically contains a large amount of
detector-related noise and the measurement unavoidably destroys
the coherence present in the qubit's initial state, it might seem
that this type of measurement cannot be used to determine the
initial state of the qubit, i.e.~at the time that the experimenter
decides to perform the measurement. Indeed, there have been a
number of studies analyzing the measurement-induced decoherence
and the evolution of the qubit's state in this situation
\cite{Gurvitz,Korotkov,Makhlin,Goan,Pilgram}, but not the question
of how to take the measurement data and extract from it
information about the initial state of the qubit. This question is
a key issue for qubit-state readout and is the main subject of
this paper. We shall show below that high-fidelity information can
be extracted from the measurement data, provided that additional
decoherence mechanisms are weak and the readout signal can be
monitored at a sufficiently short timescale. It turns out that not
only the measurement result, but also the measurement basis is
determined stochastically in this case. In spite of the
uncontrollability of the measurement basis, the measurement
results correspond properly to the initial state of the qubit. As
an example, we show how they can be used to perform quantum state
tomography on the qubit. These results show that under suitable
conditions and by proper analysis of the measurement data useful
information about the state of the qubit can be obtained.

\section{Model}

\begin{figure}[h]
\includegraphics[width=8.0cm]{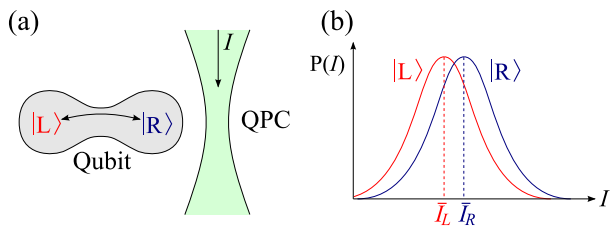}
\caption{(color online) Schematic diagrams of (a) a charge qubit
measured by a quantum point contact (QPC) and (b) the probability
distributions of the possible QPC current values for the two
charge states of the qubit. The finite widths of the probability
distributions are a result of the finite measurement time. When
the distance between the two center points in Fig.~1(b)
$|\overline{I}_R-\overline{I}_L|$ is much smaller than the widths
of the distributions, the QPC performs a weak measurement on the
qubit in the short interval under consideration. In plotting (b)
we have assumed that $\overline{I}_R>\overline{I}_L$, which would
be the case if the qubit is defined by an extra positive charge
(e.g.~a hole) tunneling between the two wells. Taking the opposite
case, i.e.~$\overline{I}_L>\overline{I}_R$, would not have any
effect on the analysis of this paper.}
\end{figure}

We consider a system composed of a charge qubit capacitively
coupled to a QPC, as illustrated in Fig.~1. The qubit can be
viewed as a system where a charged particle is trapped in a double
well potential and can occupy, and tunnel between, the localized
ground states of the two wells. We shall denote these states by
$\ket{L}$ and $\ket{R}$.

During the measurement a voltage is applied to the QPC, and a
current flows through it. We assume that the QPC measures the
qubit in the $\{\ket{L},\ket{R}\}$ basis; the current through the
QPC depends on whether the qubit is in the state $\ket{L}$ or
$\ket{R}$. We further assume that the QPC does not induce any
decoherence in the qubit's state except that associated with the
measurement-induced projection. For purposes of fully
characterizing the operation of the QPC as a detector for the
state of the qubit, it is convenient to make a few statements
about the QPC's operation when the qubit Hamiltonian is diagonal
in the charge basis and the qubit is initialized in one of the
states of the charge basis. In this case, there is no mechanism by
which the states $\ket{L}$ and $\ket{R}$ mix during the system
dynamics. If the qubit is initially in the state $\ket{L}$, the
long-time-averaged QPC current is given by $\overline{I}_L$, and
the qubit remains in the state $\ket{L}$. A similar statement
applies to the state $\ket{R}$ of the qubit, with corresponding
QPC current $\overline{I}_R$. The QPC current therefore serves as
an indicator of the qubit's state in the charge basis
$\{\ket{L},\ket{R}\}$, as long as the qubit Hamiltonian does not
mix the states of this basis.

On any finite timescale, there will be fluctuations in the QPC
current, and the observed value might deviate from
$\overline{I}_L$ or $\overline{I}_R$. The longer the period over
which the averaging is made, the smaller the fluctuations. One can
therefore define a measurement timescale that determines how long
one needs to wait in order to distinguish between the states
$\ket{L}$ and $\ket{R}$. The relation between this timescale and
the qubit's Hamiltonian-induced precession period separates two
measurement regimes: fast versus slow measurement, or
alternatively strong versus weak qubit-detector coupling. Note
that this distinction is irrelevant when the qubit Hamiltonian is
diagonal in the charge basis, since there is no mixing between the
states $\ket{L}$ and $\ket{R}$ in this case.

For the remainder of this paper, we analyze the general case where
the qubit Hamiltonian is not necessarily diagonal in the charge
basis. We shall use the basis in which the qubit Hamiltonian is
diagonal, thus
\begin{equation}
\hat{H}_{\rm q} = - E \hat{\sigma}_z /2,
\end{equation}
where $E$ is the energy splitting between the qubit's two energy
levels, and $\hat{\sigma}_z$ is the $z$-axis Pauli matrix. We
shall denote the ground and excited states of the Hamiltonian by
$\ket{0}$ and $\ket{1}$, respectively. The states of the charge
basis can be expressed as
\begin{eqnarray}
\ket{R} = \cos\frac{\beta}{2} \ket{0} + \sin\frac{\beta}{2}
\ket{1} \nonumber
\\
\ket{L} = \sin\frac{\beta}{2} \ket{0} - \cos\frac{\beta}{2}
\ket{1},
\end{eqnarray}
where $\beta$ represents the angle between the charge basis and
the energy eigenbasis.

\section{Measurement- and Hamiltonian-induced dynamics}

We start our analysis by considering a short time interval between
times $t$ and $t+\delta t$. We assume that during this time
interval a large number of electrons tunnel through the QPC, such
that it is natural to define a QPC current $I(t)$ during this
short interval. We also assume that a weak-measurement regime
exists for a properly chosen value of $\delta t$, which means that
the QPC-current probability densities (for the states $\ket{L}$
and $\ket{R}$) are broad and almost completely overlap, as shown
in Fig.~1(b). For definiteness we shall take these probability
densities to be time-independent, Gaussian functions. The interval
$delta t$ is taken to be much longer than the coherence time of
the QPC, such that the QPC's operation during this interval is
independent of the QPC's state at earlier times. Finally, we take
$delta t$ to be much shorter than the precession period of the
qubit.

We now construct matrices (or propagators) that describe the
qubit-state evolution depending on the observed QPC current
$I(t)$: when a given value of $I(t)$ is observed in the QPC, the
quantum state of the qubit is projected (possibly partially)
according to the observed value. Neglecting decoherence that is
not associated with the measurement, the projection of the qubit's
state is described by a $2 \times 2$ matrix that we shall call
$\hat{U}_M[I,\delta I,\delta t]$, where $\delta I$ defines the
size of a finite interval of QPC currents that we identify with a
single value:
\begin{equation}
\rho_{\rm q}(t+\delta t) \propto \hat{U}_M[I,\delta I,\delta t] \;
\rho_{\rm q}(t) \; \hat{U}_M^{\dagger}[I,\delta I,\delta t],
\label{Eq:Projection}
\end{equation}
where $\rho_{\rm q}$ denotes the qubit's density matrix, and
$\dagger$ represents the complex conjugate transpose. One could
say that with the introduction of $\delta I$ we are turning the
probability distributions in Fig.~1(b) into histograms with
discrete possible outcomes (This discretization will also be used
in our numerical calculations below). In order to identify the
appropriate form for $\hat{U}_M[I,\delta I,\delta t]$, we note
that the probability of obtaining the corresponding outcome is
\begin{equation}
P[I,\delta I,\delta t] = {\rm Tr} \{ \hat{U}_M^{\dagger}[I,\delta
I,\delta t] \hat{U}_M[I,\delta I,\delta t] \rho_{\rm q}(t) \}.
\label{Eq:Probability}
\end{equation}
Let us denote by $P_j[I,\delta I,\delta t]$ the probability that
the value $I$ (up to the dicretization parameters $\delta I$ and
$\delta t$) of the QPC current is observed given that the qubit is
in state $j$. We now find that the simplest, and in some sense
ideal, matrix $\hat{U}_M[I,\delta I,\delta t]$ that satisfies
Eq.~(\ref{Eq:Probability}) is given by
\begin{eqnarray}
\hat{U}_M[I,\delta I,\delta t] & = &
\sqrt{P_L[I,\delta I,\delta t]} \ket{L}\bra{L} + \nonumber \\
& & \hspace{0cm} \sqrt{P_R[I,\delta I,\delta t]} \ket{R}\bra{R}.
\label{Eq:UM}
\end{eqnarray}
This matrix could be followed by a unitary transformation that
does not affect Eq.~(\ref{Eq:Probability}). Any such
transformation can be incorporated into the analysis
straightforwardly.

In addition to the measurement-induced evolution described by
Eq.~(\ref{Eq:UM}), the qubit Hamiltonian induces a unitary
evolution in the qubit's state during the time interval $t$ to
$t+\delta t$: taking $\hbar=1$
\begin{eqnarray}
\hat{U}_H [\delta t] = \exp\left\{-i \hat{H}_{\rm q} \delta t
\right\} \approx 1 + i \frac{E}{2} \delta t \hat{\sigma}_z.
\label{Eq:UH}
\end{eqnarray}

The matrices $\hat{U}_M[I,\delta I,\delta t]$ and $\hat{U}_H
[\delta t]$ can now be combined to give the total evolution matrix
\begin{eqnarray}
\hat{U}[I(t),\delta I,\delta t] & = & \hat{U}_M[I(t),\delta
I,\delta t] \times \hat{U}_H [\delta t].
\label{Eq:U_delta_t}
\end{eqnarray}
Note that both $\hat{U}_M[I,\delta I,\delta t]$ and $\hat{U}_H
[\delta t]$ are approximately proportional to the unit matrix in
the limit $\delta t \rightarrow 0$, with lowest-order corrections
of order $\delta t$. The operators $\hat{U}_M[I,\delta I,\delta
t]$ and $\hat{U}_H [\delta t]$ therefore commute to first order in
$\delta t$.

When a given QPC output signal $I(t)$ [from the initial time $t=0$
until $t=t_f$] is observed, one can take the corresponding
short-time evolution matrices explained above and use them to
construct the total evolution matrix $\hat{U}_{\rm
Total}[I(t:0\rightarrow t_f),\delta I,\delta t]$. Using the unit
matrix as the total evolution matrix for $t=0$, we find that
\begin{eqnarray}
\hat{U}_{\rm Total}[I(t:0\rightarrow t_f),\delta I,\delta t] & = &
\hat{U}[I(t_f-\delta t),\delta I,\delta t] \times \cdots \times
\nonumber \\ & & \hat{U}[I(0),\delta I,\delta t].
\end{eqnarray}

Once the $2 \times 2$ matrix $\hat{U}_{\rm Total}[I(t:0\rightarrow
t_f),\delta I,\delta t]$ is calculated, one can divide it into two
parts, a measurement matrix $\hat{U}_{\rm Meas}[I(t:0\rightarrow
t_f),\delta I,\delta t]$ followed by a unitary transformation
$\hat{U}_{\rm Rot}[I(t:0\rightarrow t_f),\delta I,\delta t]$:
\begin{eqnarray}
& & \hat{U}_{\rm Total}[I(t:0\rightarrow t_f),\delta I,\delta t] =
\nonumber
\\ & & \hspace{0.3cm} \hat{U}_{\rm Rot}[I(t:0\rightarrow t_f),\delta I,\delta t] \times
\hat{U}_{\rm Meas}[I(t:0\rightarrow t_f),\delta I,\delta t].
\end{eqnarray}
%
The matrix $\hat{U}_{\rm Meas}$ has the form
\begin{equation}
\hat{U}_{\rm Meas} = \sqrt{P_1} \ket{\psi_1}\bra{\psi_1} +
\sqrt{P_2} \ket{\psi_2}\bra{\psi_2},
\end{equation}
where $\ket{\psi_1}$ and $\ket{\psi_2}$
are two orthogonal states. The states $\ket{\psi_1}$ and
$\ket{\psi_2}$ represent the measurement basis that corresponds to
the output signal $I(t)$, and the parameters $P_i$ are the
probabilities that the outcome defined by $I(t)$, $\delta I$ and
$\delta t$ is obtained given that the qubit was initially in the
state $\ket{\psi_i}$. With a simple calculation, one can see that
the measurement fidelity is given by (see
e.g.~\cite{AshhabMeasurement1})
\begin{equation}
{\rm Fidelity} = \left| \frac{P_1-P_2}{P_1+P_2} \right|.
\end{equation}

To summarize, the QPC output-current signal can be used to derive
the matrix $\hat{U}_{\rm Total}[I(t:0\rightarrow t_f),\delta
I,\delta t]$. This matrix can then be used to determine the
measurement basis, the measurement result (i.e.~$\pm 1$ along the
measurement axis), the fidelity (or in other words the degree of
certainty about the obtained measurement result) and the final
state of the qubit (given by the measurement result transformed by
the unitary, i.e.~rotation, matrix $\hat{U}_{\rm
Rot}[I(t:0\rightarrow t_f),\delta I,\delta t]$). Note that when
the measurement fidelity approaches one, the final state is a pure
state that can be determined even without any {\it a priori}
knowledge about the initial state.

\section{Results and discussion}

We now present the results of our numerical calculations. The
calculations were performed by analyzing a sequence of discrete
events, with each event representing a time steps of size $\delta
t$. We also discretize the possible values of QPC current. We have
verified that the results presented below are insensitive to the
exact discretization parameters, as long as we take $E\delta t \ll
1$ and there are a large number of possible QPC current values.
The qubit is initialized in a given state that depends on the
specific calculation. In each time step, the value of the QPC
current is determined stochastically using
Eq.~(\ref{Eq:Probability}). Based on the obtained value, the
qubit's state evolves as described in Eq.~(\ref{Eq:Projection}).
Following this weak-measurement step, a unitary transformation of
the form of Eq.~(\ref{Eq:UH}) is applied to the qubit's state.
After a sufficiently long QPC signal is obtained, this signal (in
all its details) is taken and used to extract the measurement
matrix $\hat{U}_{\rm Meas}$ explained above. This matrix is then
used to extract the measurement basis and fidelity.

The strength of the qubit-QPC coupling is determined by the
relation between two parameters in the numerical calculations: (1)
the width, or standard deviation $\sigma$, of the QPC-current
distribution functions and (2) the distance between the average
values of these distribution functions
$(\overline{I}_R-\overline{I}_L)$. It is more convenient, however,
to present the results in terms of a different parameter that
characterizes the qubit-QPC coupling strength, namely
$E\tau_m/(2\pi)$, where $\tau_m$ is the timescale needed to obtain
sufficient QPC signal to read out the state of the qubit (for the
time being one can think of this definition as applying to the
case when $\beta=0$; but see below). If one considers $N$ of the
small steps considered above, the standard deviation of the QPC's
averaged signal scales as $\sigma/\sqrt{N}$ (note that $\sigma$ is
the standard deviation for one step). The measurement time
$\tau_m$ can now be naturally defined as the product of the time
step $\delta t$ and the value of $N$ at which
$2\sigma/\sqrt{N}=|\overline{I}_R-\overline{I}_L|$. The
measurement time is therefore given by
\begin{equation}
\tau_m = \frac{4\sigma^2 \delta
t}{|\overline{I}_R-\overline{I}_L|^2}.
\end{equation}

\begin{figure}[h]
\includegraphics[width=8.0cm]{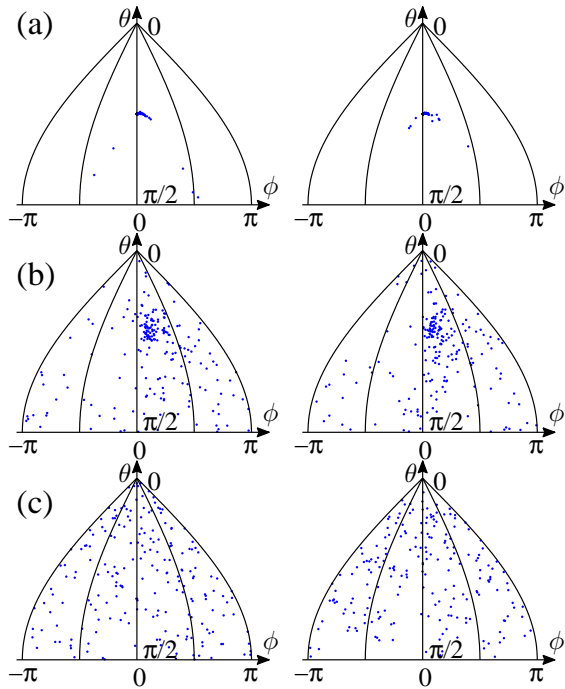}
\caption{(color online) The spherical coordinates $\theta$ and
$\phi$ that define the stochastically determined measurement bases
obtained in simulations of the experiment under consideration
[note that $(\theta,\phi)=(0,0)$ for the energy eigenbasis and
$(\theta,\phi)=(\beta,0)$ for the charge basis]. Each figure
contains 200 points. In Fig.~2(a) $E\tau_m/(2\pi)=0.01$, i.e.~deep
in the strong-coupling regime. In Fig.~2(b) $E\tau_m/(2\pi)=0.2$
[intermediate-coupling regime], and in Fig.~2(c)
$E\tau_m/(2\pi)=5$ [weak-coupling regime]. In all the figures,
$\beta=\pi/4$. In each set, the figure on the left is generated
using the initial state $\ket{L}$, and the one on the right is
generated using the initial state $\ket{0}$. Each set is an
identical pair, up to statistical fluctuations, demonstrating that
the initial state plays no role in determining the measurement
basis.}
\end{figure}

First, in Fig.~2 we show the spherical coordinates of the
(stochastically determined) measurement bases for different levels
of qubit-detector coupling. In the strong-coupling,
fast-measurement regime (Fig.~2a), the measurement basis is always
the charge basis, which is the natural measurement basis for the
detector under consideration. As the qubit-detector coupling
strength is reduced (Fig.~2b), the measurement bases start to
deviate from the charge basis, and they develop some statistical
spread. This region could be called the intermediate-coupling
regime. In the weak-coupling, slow-measurement regime (Fig.~2c),
the measurement bases are spread over all the possible directions.

\begin{figure}[h]
\includegraphics[width=6.0cm]{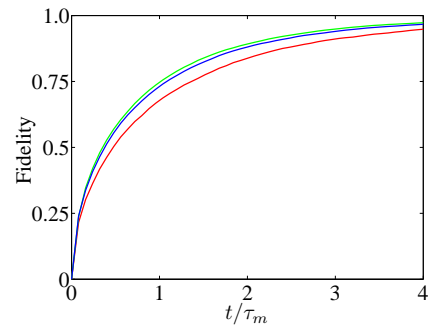}
\caption{(color online) The measurement fidelity as a function of
measurement duration for three different values of the angle
$\beta$ between the charge basis and the energy eigenbasis:
$\beta=0$ (red; lowest line), $\pi/4$ (green) and $\pi/2$ (blue).
Here $E\tau_m/(2\pi)=5$, i.e.~deep in the weak-coupling regime.
The fidelity increases as the measurement duration increases, but
the fidelity is essentially independent of the angle $\beta$.}
\end{figure}

In Fig.~3, we plot the measurement fidelity as a function of
measurement duration for three different values of $\beta$,
keeping all other parameters fixed. We can see that the fidelity
approaches one for long enough measurement duration, regardless of
the angle $\beta$. In fact, the fidelity is almost independent of
$\beta$. This result shows that even though more complicated
analysis is needed to extract useful measurement information when
$\beta \neq 0$, the information acquisition rate is only slightly
affected by the coherent, Hamiltonian-induced precession.

The fact that the measurement basis is generally unpredictable,
and therefore uncontrollable, is a rather strange phenomenon from
a fundamental point of view. From a practical point of view, one
can wonder whether anything useful can be done with such
measurements that are performed in a stochastically determined
basis. If one absolutely requires a measurement in a given basis,
measurement results in different bases would be less useful. One
could then treat the deviation of the observed measurement basis
from the desired one as an experimental error and deal with it
accordingly.

In the above discussion, we have ignored any decoherence other
than that associated with the measurement-induced projection. If
the measurement time $\tau_m$ is much smaller than the timescale
of decoherence caused by other mechanisms, the measurement is
completed with minimal effects of any additional decoherence. Our
results are valid in this case. A number of different types of
detectors, including QPCs, are approaching this limit where the
decoherence is limited by the measurement \cite{Clerk}, indicating
that our results could be observable in such systems.


{\it Quantum state tomography.---} One example of a procedure
where the uncontrollability of the measurement basis can be
harmless is quantum state tomography (QST). In QST, one is given a
large number of copies of the same quantum state, and the goal is
to deduce this state. Typically, one measures the operator
$\hat{\sigma}_x$ for one third of the copies, and similarly for
$\hat{\sigma}_y$ and $\hat{\sigma}_z$. Once the average values
$\left\langle\hat{\sigma}_x\right\rangle$,
$\left\langle\hat{\sigma}_y\right\rangle$ and
$\left\langle\hat{\sigma}_z\right\rangle$ are known, the density
matrix is reconstructed straightforwardly:
\begin{equation}
\rho = \frac{1}{2} \left( 1 +
\left\langle\hat{\sigma}_x\right\rangle \hat{\sigma}_x +
\left\langle\hat{\sigma}_y\right\rangle \hat{\sigma}_y +
\left\langle\hat{\sigma}_z\right\rangle \hat{\sigma}_z \right).
\end{equation}

In the present case, the measurement bases are chosen
stochastically, and in principle no two of them are the same. One
must therefore reconstruct the unknown quantum state using a
procedure that allows for data taken from measurements made in any
combination of bases. One such procedure is the minimization of
the function
\begin{equation}
{\cal T}(r,\theta,\phi) = \sum_j \left[ 1 - r \cos
\Omega(\theta,\phi,\theta_j,\phi_j) \right]^2,
\end{equation}
where $r$, $\theta$ and $\phi$ are the spherical coordinates of a
point in the Bloch sphere; $j$ is an index labelling the different
runs of the experiment; the direction ($\theta_j,\phi_j$) defines
the qubit state obtained in a given measurement; and
$\Omega(\theta,\phi,\theta_j,\phi_j)$ is the angle between the
directions $(\theta,\phi)$ and $(\theta_j,\phi_j)$. Assuming that
the measurement bases cover all possible directions (see
e.g.~Fig.~2c), the convergence of this procedure to the correct
density matrix should be similar to the convergence of the
standard QST procedure explained in the previous paragraph.

We have simulated QST by repeating the measurement procedure a
large number of times, obtaining a set of measurement results (in
the form of pre-measurement qubit states), and then minimizing the
function ${\cal T}(r,\theta,\phi)$ with the respect to $r$,
$\theta$ and $\phi$. We have chosen several initial states
covering the Bloch sphere, and the tomography procedure
consistently produced the initial state of the qubit for the
parameters of Figs.~2(b,c). For strong qubit-detector coupling
[see Fig.~2(a)], the procedure becomes unreliable, because the
vast majority of the measurements are performed in one basis.

\section{Conclusion}

In conclusion, we have analyzed the question of what information
can be extracted from the output signal of a QPC that weakly
probes the charge state of a charge qubit when the qubit
Hamiltonian induces oscillations between the different charge
states. We have shown that the measurement basis is determined
stochastically every time the measurement is repeated. In the case
of weak qubit-detector coupling, the possible measurement bases
cover all the possible directions. The measurement basis and
result can both be extracted from the QPC output signal. We have
also shown that the information acquisition rate is independent of
the angle $\beta$ between the direction defining the charge basis
and that defining the qubit Hamiltonian. In other words, given
enough time, the detector will produce a high-fidelity measurement
result, regardless of the value of $\beta$. These results show
that, under suitable conditions and by proper analysis, the
detector's ability to obtain high-fidelity information about the
state of the qubit is not affected by the competition between the
measurement and coherent-precession dynamics. More detailed
analysis of the results discussed in this paper is presented
elsewhere \cite{AshhabLongVersion}.

We would like to thank A. J. Leggett for useful discussions. This
work was supported in part by the NSA, LPS, ARO, and NSF grant
No.~EIA-0130383. J.Q.Y. was also supported by the ``973" Program
grant No. 2009CB929300, the NSFC grant No. 10625416, and the MOST
grant No. 2008DFA01930.


\begin{thebibliography}{99}

\bibitem{SolidStateReviews} See, e.g., J. Q. You and F. Nori, Phys. Today
{\bf 58} (11), 42 (2005); G. Wendin and V. Shumeiko, in {\it
Handbook of Theoretical and Computational Nanotechnology}, ed. M.
Rieth and W. Schommers (ASP, Los Angeles, 2006); J. Clarke and F.
K. Wilhelm, Nature {\bf 453}, 1031 (2008); D. Loss and D. P.
DiVincenzo, Phys. Rev. A {\bf 57}, 120 (1998).

\bibitem{Gurvitz} S. A. Gurvitz, Phys. Rev. B {\bf 56}, 15215
(1997); B. Elattari and S. A. Gurvitz, Phys. Rev. Lett. {\bf 84},
2047 (2000).

\bibitem{Korotkov} A. N. Korotkov, Phys. Rev. B {\bf 60}, 5737 (1999);
A. N. Korotkov and D. V. Averin, {\it ibid.} {\bf 64}, 165310
(2001).

\bibitem{Makhlin} Y. Makhlin, G. Sch\"on, and A. Shnirman, Phys. Rev.
Lett. {\bf 85}, 4578 (2000).

\bibitem{Goan} H.-S. Goan, G. J. Milburn, H. M. Wiseman, and H. B. Sun, Phys.
Rev. B {\bf 63}, 125326 (2001).

\bibitem{Pilgram} S. Pilgram and M. B\"uttiker, Phys. Rev. Lett. {\bf 89},
200401 (2002).

\bibitem{QuantumCapacitance} We shall not consider measurement
methods based on probing the capacitance of the qubit; M. A.
Sillanp\"a\"a, T. Lehtinen, A. Paila, Yu. Makhlin, L. Roschier,
and P. J. Hakonen, Phys. Rev. Lett. {\bf 95}, 206806 (2005);  T.
Duty, G. Johansson, K. Bladh, D. Gunnarsson, C. Wilson, and P.
Delsing, Phys. Rev. Lett. {\bf 95}, 206807 (2005).

\bibitem{Johansson} For a review on different readout methods in
solid-state qubits, see e.g. G. Johansson, L. Tornberg, V. S.
Shumeiko, and G. Wendin, J. Phys.: Condensed Matter {\bf 18}, S901
(2006).

\bibitem{AshhabCoupling} Interestingly, weak measurement can have
the advantage of being insensitive to undesirable mixing or
`contamination' between the states of the measurement basis; S.
Ashhab, A. O. Niskanen, K. Harrabi, Y. Nakamura, T. Picot, P. C.
de Groot, C. J. P. M. Harmans, J. E. Mooij, and F. Nori, Phys.
Rev. B {\bf 77}, 014510 (2008); L. Fedichkin, M. Shapiro, and M.
I. Dykman, Phys. Rev. A {\bf 80}, 012114 (2009).

\bibitem{AshhabMeasurement1} S. Ashhab, J. Q. You, and F. Nori, Phys.
Rev. A {\bf 79}, 032317 (2009).

\bibitem{Clerk} A. A. Clerk, M. H. Devoret, S. M. Girvin, F. Marquardt,
R. J. Schoelkopf, Rev. Mod. Phys. (in press); G. M. Reuther, D.
Zueco, P. H\"anggi, and S. Kohler, Phys. Rev. Lett. {\bf 102},
033602 (2009).

\bibitem{AshhabLongVersion} S. Ashhab, J. Q. You, and F. Nori, New
J. Phys. {\bf 11}, 083017 (2009).

\end{thebibliography}
\end{document}